\documentclass[pre,twocolumn,showpacs,amsmath,amssymb]{revtex4}

\usepackage{graphicx}
\usepackage{dcolumn}
\usepackage{bm}
\usepackage{epstopdf}

\begin{document}


\title{Phase Transitions in Operational Risk}

\author{Kartik Anand}
\email{kartik.anand@kcl.ac.uk}

\author{Reimer K\"uhn}
 \email{reimer.kuehn@kcl.ac.uk}

\affiliation{Department of Mathematics, King's College London, Strand, London WC2R 2LS, UK}

\date{\today}

\begin{abstract}
In this paper we explore the functional correlation approach to operational risk. We consider networks with heterogeneous a-priori conditional and unconditional failure probability. In the limit of sparse connectivity, self-consistent expressions for the dynamical evolution of order parameters are obtained. Under equilibrium conditions, expressions for the stationary states are also obtained. The consequences of the analytical theory developed are analyzed using phase diagrams. We find co-existence of operational and non-operational phases, much as in liquid-gas systems. Such systems are susceptible to discontinuous phase transitions from the operational to non-operational phase via catastrophic breakdown. We find this feature to be robust against variation of the microscopic modelling assumptions.
\end{abstract}



\maketitle

\section{\label{sec: Introduction}Introduction}

Management and mitigation of risk events are major concerns for banks. The goal is two-fold: first, quantitatively assess the risk in terms of potential financial loss and second, develop solutions to control and buffer the impact of these losses. To facilitate systematic analysis, risk events are broadly classified into three categories: (i) Market Risk (MR), (ii) Credit Risk (CR) and (iii) Operational Risk (OR). MR refers to fluctuations in stock indices, changes in interest rates, foreign exchange parities or commodity (e.g., gold, oil, etc) prices. CR refers to loan defaults when companies go bankrupt. Research on understanding risk and developing sophisticated models has traditionally focused on MR and CR, while OR was initially subsumed under ``other", non credit or market risks. Subsequent spectacular catastrophes including the bankruptcy of the Orange County municipality, California, USA in 1994 \cite{Artzner:1999} and the collapse of Baring Investment Bank, London, United Kingdom in 1995 \cite{Marrison:2002}, which were neither attributed to MRs nor CRs, helped establish OR as a risk category of its own. The Basel Committee for Banking Supervision (BCBS), an international regulatory body, now stipulates that banks must explicitly reserve a portion of equity capital against OR.

Note that the above listed three main risk categories are not intended to be exhaustive. Other risk categories exist (e.g., liquidity risk is an important category for bank management, or fiduciary and compliance risks, which arise from the judicial responsibilities of the banks' customers). Depending on circumstances, these may indeed outweigh the importance of the three `main' risk categories mentioned above.

The Basel II document \cite{BaselII:2005}, released by BCBS in 2001 and revised in 2005, is a guideline on banking regulation. Under Basel II, OR is defined as the risk of losses resulting from inadequate or failed internal processes, people and systems, or from external events. From a practical and management perspective it is reasonable to categorize ORs as events based on causes and specific effects. Possible categorizations, as described in \cite{Rolfes:2001}, are: (i) human processing errors, (ii) human decision errors, (iii) system (software or hardware) errors, (iv) process design errors, (v) fraud and theft and (vi) external damages.
 
To quantify the capital that must be allocated for operational failures, Basel II suggests three methodologies: (i) Basic Indicator Approach (BIA), (ii) Standardized Approach (SA) and (iii) Advanced Measurement Approach (AMA). Under BIA, the required capital is determined by taking 15\% of the banks' average gross income over the previous three years. The SA is only slightly more advanced in that the banks' operations are divided into 8 business lines, each of which has a specific weight. The required capital is calculated as the weighted average of the (non-negative) gross income from the business lines over the previous three years. Under AMA, banks are responsible for designing their own measurement approach and setting assumptions on the loss distribution. However, banks must demonstrate that their approach captures potentially severe ``tail" loss events. The use of external and internal loss data as well as internal expertise is permitted in the evaluation.

Under AMA, the Loss Distribution Approach (LDA) is a sophisticated and popular measure. Herein the required capital is determined by the Value-at-Risk (VaR) \cite{Marrison:2002} over all OR categories. VaR is defined, over a specified risk horizon $T$ as the loss not exceeded with probability $q$, in excess of the expected loss that can happen under normal economic conditions. The VaR is thus dependent on the nature of the loss frequency and severity distributions.

Choices for the loss severity distribution function include log-normal, Gamma, Beta and Weibull distributions. A Poisson or negative binomial distribution is often used for the loss frequency distribution. The common approach to estimate the loss distribution is to first assume that OR categories are independent. Subsequently, for each category $i$, one draws a realization $N_i$ from the loss frequency distribution and samples $N_i$ realizations of the loss severity $X_i^m\,(m=1,\ldots,N_i)$. The loss is then calculated as
\begin{equation}
\label{eq:loss per process}
L_i\,=\,\sum_{m=1}^{N_i}X_i^m\,.
\end{equation}
Finally, drawing a histogram of outcomes of normalized $L_i$ one obtains the loss distribution for each category. The capital to be allocated is the VaR of the sum of losses over all categories, given a time horizon $T$ and confidence level $q$.

Subsequent developments in modelling OR have focused on incorporating more realistic loss frequency and severity distributions. In \cite{Chernobai:2004} Chernobai and Rachev provide evidence in favor of using Stable Paretian Distributions. Similarly, in \cite{Allen:2004} Allen et. al., incorporate a Generalized Pareto Distribution.

A critical point concerning LDA and subsequent analytical development is the assumption of independence between various OR categories. However, a moment's reflection will lead us to realize that these categories, which may be conceptualized as sets of processes, are functionally dependent.

To put this in context we illustrate it with the collapse of Baring Investment Bank. A trader in the banks' Singapore office hid trading losses by forging official documents (trading decision errors resulting in fraud), thereby allowing discrepancies to go un-noticed by higher officials. Pleased with his apparent performance, the bank made a mistake of retaining him as Chief Trader (process design error in light of the bank not having adequate checks and balances, which led to decision errors by the banks' managers). In December 1994, compounded by increasing losses and in order to overcome the predicament, the trader bet on the Tokyo Stock Exchange, assuming that the index would not fall below 19,000. However, in January 1995 a devastating earthquake hit the Japanese city of Kobe, resulting in the index to plummet by 7\% in one week, and below 19,000 (series of decision errors compounded by external damage). By the time the bank discovered what had happened, the trader had lost \$1.3 billion, effectively bankrupting the bank.

This example shows that the assumption of independence between processes is not realistic. Recent efforts have therefore been directed to capture the influence of interactions between processes. A model based on the lattice gas analogy was proposed in \cite{Kuhn:2003}. Investigations were conducted via Monte Carlo (MC) simulations and using a mean-field approximation for a homogeneous and fully connected process network. The model parameters were shown to be related to both conditional and unconditional failure probabilities. Furthermore, avalanches of process failures were shown to be possible through bubble nucleation as in first order phase transition. This model was subsequently elaborated by Leipold and Vanini \cite{Leippold:2005}. In \cite{Clemente:2004}, Clemente and Ramano present a case study, substantiated by MC simulations, incorporating realistic dependences between processes.

At this point, it is important to note that the concept of OR is not restricted to the banking industry, but is also systemic to any large economy or commerce. Over the past few decades, markets have been subject to considerable de-regulation and globalization. These forces, coupled with an increasing reliance on sophisticated information technology have allowed businesses to develop more efficient operational practices, which include Business Processes Outsourcing (BPO) and automated data collection, storage and retrieval techniques. These advances have led to increased mutual dependencies between economic processes, resulting in a heightened susceptibility for catastrophic breakdowns and thus significant financial losses. A through understanding of the dynamics of interacting process networks is more than ever desirable.

Therefore, in consideration of the above, the aim of the present paper is twofold: (i) provide detailed analytical underpinning for the main findings of \cite{Kuhn:2003}, and (ii) in doing so, broaden the scope of that investigation to highlight the fact that the possibility of catastrophic breakdown in networks of interacting processes is a robust phenomenon under variation of the underlying model assumptions.

The remainder of the paper is organized as follows. In Sec. \ref{sec:Model Definition} we introduce our model. Sec. \ref{sec:Model sols} begins by introducing three distinct techniques, covering non-equilibrium dynamics and equilibrium statistical mechanics, which we use to derive dynamical evolution and stationary state solutions. In Sec. \ref{sec: Heuristic Solution} we detail a heuristic solution, valid in the case of asymmetric, uncorrelated interactions between processes.  In Sec. \ref{sec: GFA solution} a more systematic study, based on a Generating Functional Analysis (GFA) and valid for an arbitrary degree of interaction symmetry is provided. In Sec. \ref{sec: stationary states solution} we investigate the stationary behavior of a fully symmetric network using techniques from equilibrium statistical mechanics. In Sec. \ref{sec: Results} we produce phase diagrams for process networks, exhibiting regions in parameter space where operational phases coexist with non-operational phases. We subsequently evaluate a loss distribution and thereby extend the scope of LDA to account for correlations between processes. Finally, in Sec. \ref{sec: Conclusion} we provide concluding remarks and describe possible extensions for further work. Some of the more technical details from GFA and analysis of stationary states are relegated to appendices \ref{apx: GFA} and \ref{apx: Replica}, respectively.

\section{\label{sec:Model Definition} Model Definitions}
In this section we describe the statistical model of interacting processes, which was previously introduced in \cite{Kuhn:2003}. Each process $i$ ($i\,=\,1,\,\ldots\,N$) is defined by its state at time $t$, which is referred to as $\eta_i(t)$. A two-state model is considered: $\eta_i(t)\,\in\,\{0,1\}$. At time $t$ a process can be either up and running, $\eta_i(t)\,=\,0$, or broken down, $\eta_i(t)\,=\,1$. In order to maintain a stable running state over the time increment $t\,\to\,t + 1$, each process $i$ needs to receive support at time $t$, which is denoted $U_i(t)$ and takes the form
\begin{equation}
\label{eq:support for process}
U_i(t) \,=\, \vartheta_i\,-\,\sum_{j} J_{ij}\,\eta_j(t)\,-\,\xi_i(t)\,.
\end{equation}
Here, $\vartheta_i\,\in\,\mathbb{R}$ denotes the average support received by a process in a fully functional environment, while $J_{ij}\,\in\,\mathbb{R}$ represents the impact on the average support if process $j$ breaks down. Finally, the $\xi_i(t)$ are zero mean, Gaussian random fluctuations, which represent non-systematic internal and external perturbations to the environment (e.g., fire, earthquake, voltage fluctuations within electric supplies, etc.).

A process breaks down in the next time step if $U_i(t)\,<\,0$. Thus, the dynamics of a processes' state is given by
\begin{equation}
\label{eq:process state}
\eta_i(t+1)\,=\,\Theta \left( \sum_{j}J_{ij}\,\eta_j(t)\,-\,\vartheta_i\,+\,\xi_i(t) \right)\,,
\end{equation}
where $\Theta(\ldots)$ represents the Heaviside function. The time-step, is taken to represent one day.

By suitable rescaling of the average support and impact parameters $\vartheta_i$ and $J_{ij}$, respectively, the $\xi_i(t)$ can be taken to have unit variance. The rescaled variables, can then be related to the unconditional and conditional failure probabilities. Consider a situation wherein all processes are working at time-step $t$. The probability that process $i$ will break down in the next step, referred to as $p_i$, is given by integrating Eq. (\ref{eq:process state}) over the noise $\xi_i(t)$. Thus
\begin{equation}
\label{eq: unconditional failure prob}
p_i\,=\,\Phi \left( -\vartheta_i \right)\,, 
\end{equation}
where $\Phi(x)\,=\,\frac{1}{2}\,\left(1\,+\,{\rm erf}\left(\frac{x}{\sqrt{2}}\right)\right)$. Similarly, defining $p_{i|j}$ as the probability process $i$ will break down in the next step, given that currently process $j$ is broken down while all others are working, gives us
\begin{equation}
\label{eq: conditional failure probs}
p_{i|j}\,=\,\Phi \left( J_{ij}\,-\,\vartheta_i \right)\,.
\end{equation}
These relations may be inverted to obtain expressions for the model parameters in terms of the failure probabilities,
\begin{eqnarray}
\label{eq:model parameters}
\vartheta_i\,=\,-\,\Phi^{-1} (p_i)\,, \qquad J_{ij}\,=\, \Phi^{-1} (p_{i|j})\,-\,\Phi^{-1} (p_i)\,.
\end{eqnarray}

In general, this model is not analytically tractable. In \cite{Kuhn:2003} investigations were conducted via MC simulations and using a mean-field approximation for a homogeneously connected network with uniform conditional and unconditional failure probabilities. In what follows, we study and solve the model in a more interesting regime where conditional and unconditional failure probabilities are heterogeneous across the system.

We begin by noting that if $J_{ij}\,<\,0$, the breakdown of process $j$ adds support and is beneficial to process $i$, i.e., the two processes are competing. Such a situation is undesirable, but tends to occur on a small scale in large organizations. Thus, when considering a primarily cooperative environment, it is desireable to have, with high probability, $J_{ij}\,>\,0$. 

In our framework, each process does not interact with all, but instead a fraction of the other processes. We explicitly incorporate this feature by decomposing
\begin{equation}
\label{eq:decompose bond disorder}
 J_{ij}\,=\,c_{ij}\,\tilde{J}_{ij}\,,
\end{equation}
where $c_{ij}\,\in\,\{0,1\}$ are connectivity coefficients and $\tilde{J}_{ij}\,\in\,\mathbb{R}$ describes the magnitude of impact. We assume $c_{ij}\,=\,c_{ji}$. While this is a reasonable assumption, it is important to realize that the impact magnitudes are not necessarily symmetric. For example, consider a main-frame computer connected to a dummy terminal computer. If the main-frame crashes, we cannot use the terminal. However, if the terminal computer breaks down, it is highly unlikely that it would effect the operations of the main-frame. Thus, in general, $\tilde{J}_{ij}\,\neq\,\tilde{J}_{ji}$. The connectivity coefficients are described by the following distribution
\begin{eqnarray}
\label{eq:connectivity distribution}
P(c_{ij})\,=\,\left(1\,-\,\frac{c}{N}\right)\delta_{[c_{ij},0]}\,+\,\frac{c}{N}\delta_{[c_{ij},1]}\,,
\end{eqnarray}
where $c\,\in\,\mathbb{R}^{+}$ is the average connectivity per process. In the case of finite $N$, the adjacency matrix describes an Erd\"os-R\'enyi random graph \cite{Erdos:1959}.

In what follows we consider the extreme dilution limit, wherein $N\,\to\,\infty$, $c\,\to\,\infty$, such that $c/N\,\to\,0$. This is achieved by taking $c\,=\,{\cal O}(\log(N))$. This assumption has important implications on the structure of the connectivity graph. Firstly, each node in the graph will be connected to a vanishing fraction of the total number of nodes. Secondly, the length of a loop is ${\cal O}(\log(N))$ \cite{Derrida:1987}. Thus, taking $N\,\to\,\infty$, the probability of finding loops of finite length tends to $0$. The environment about each node is locally tree-like.

The magnitude of the impact parameters $\tilde{J}_{ij}$ are taken to be quenched, i.e., fixed random quantities. To allow for the thermodynamic limit $N\,\to\,\infty$ and $c\,\to\,\infty$  the mean and variance of the $\tilde{J}_{ij}$ must scale with $c$. We put
\begin{equation}
\label{eq:define J_ij}
\tilde{J}_{ij}\,=\,\frac{J_0}{c}\,+\,\frac{J}{\sqrt{c}}x_{ij}\,,
\end{equation}
where the $x_{ij}$ are zero mean and unit variance random variables. The mean and variance of $\tilde{J}_{ij}$ are parameterized by $J_0\,\in\,\mathbb{R}$ and $J\,\in\,\mathbb{R}$ respectively. We note that if $J_0\,>\,0$ the interactions between processes are, on average, supportive, which is the regime of interest. Secondly small $J$ suppresses the probability of having of negative $\tilde{J}_{ij}$. In addition, small $J$ reduces the effects of frustration \cite{Fischer:1991}. Finally, we choose the $x_{ij}$ to be independent in pairs and have the following moments
\begin{equation}
\label{eq:define x_ij}
\overline{x_{ij}}\,=\,0\,, \qquad \overline{x_{ij}\,x_{kl}}\,=\,\delta_{[i,k]}\,\delta_{[j,l]}\,+\,\alpha\,\delta_{[i,l]}\,\delta_{[j,k]}\,.
\end{equation}
The parameter $\alpha\,\in\,[0,1]$ describes the degree of correlations between $\tilde{J}_{ij}$ and $\tilde{J}_{ji}$, with fully symmetric interactions given by $\alpha\,=\,1$.

\section{\label{sec:Model sols} Model Solutions}
Here we investigate the dynamics and the stationary states of the model introduced in Sec. \ref{sec:Model Definition}.

For uncorrelated interactions, i.e., $\alpha=0$, a solution can be obtained by a heuristic argument, following lines of reasoning previously used to study statistical mechanics of disordered systems, in particular neural networks \cite{Derrida:1987}-\cite{Kree:1992}. This solution relies on assuming weak and negligible correlations between the quenched and dynamic random variables, which cannot be easily justified in a rigorous manner.

A more exact and formal treatment of the dynamics is possible using generating functional analysis (GFA) \cite{Dominicis:1978}. This technique facilitates the evaluation of order parameters {\it ab initio} and is applicable for arbitrary degree of correlation, $\alpha$, between processes. In addition, it provides a non-trivial check to show that the heuristic solution is exact.

Finally, for fully symmetric networks, $\alpha=1$, we study the stationary states using techniques from equilibrium statistical mechanics. This requires the use of thermal instead of Gaussian noise. 

\subsection{\label{sec: Heuristic Solution} Heuristic Solution}
We begin by observing that interactions of process $i$ are described by the local field $h_i(t)\,=\,\sum_{j}c_{ij}\,\tilde{J}_{ij}\,\eta_j(t)$. The state $\eta_i(t+1)$ depends on the $\eta_j(t)$ that contribute to $h_i(t)$. Furthermore, the states $\eta_j(t)$ depend on $\eta_i(t-1)$ through $h_j(t-1)$. There is feed-back in the system, which induces correlations between the local fields at time $t$.

In the present mean-field approach we evaluate the statistics of $h_i(t)$ by appealing to the law of large numbers and central limit theorem. The applicability of these tools relies on assuming that the contributions to $h_i(t)$ are weakly correlated if not independent. The condition $\alpha\,=\,0$ together with the limit of sparse connectivity, $c/N\,\to\,0$, entail that contributions to $h_i(t)$ are uncorrelated at finite times. This allows us to describe $h_i(t)$ as a Gaussian random quantity.

Concentrating on the local field $h_i(t)$, we incorporate the definition of $J_{ij}$ given by Eq. (\ref{eq:define J_ij}) to rewrite the expression as
\begin{equation}
\label{eq:expand local field}
h_i(t) \,=\, \frac{J_0}{c}\sum_{j}c_{ij}\,\eta_j(t)\,+\,\frac{J}{\sqrt{c}}\sum_{j}c_{ij}\,x_{ij}\,\eta_j(t)\,.
\end{equation}
In the limit $N\,\to\,\infty$, the mean and variance of $h_i(t)$ are given by 
\begin{equation}
\label{eq:local field average}
\overline{\langle h_i(t) \rangle}^{c_{ij},x_{ij}}\,\simeq\,\frac{J_0}{N}\sum_j \overline{\langle \eta_j(t) \rangle}^{c_{ij},x_{ij}}\,,
\end{equation}
and 
\begin{equation}
\label{eq:local field var}
\sigma^2(h_i(t))\,\simeq\, \frac{J^2}{N}\sum_{j}\overline{\langle \eta_j(t)\rangle}^{c_{ij},x_{ij}}
\end{equation}
respectively, \cite{Derrida:1987}. The angled brackets $\langle (\ldots) \rangle$ refer to the average over noise terms, $\xi_i(t)$, while the over-bar, $\overline{(\ldots)}^{c_{ij},x_{ij}}$ refers to the average over the coupling parameters, $c_{ij}$ and $x_{ij}$. 

In deriving these results, one assumes that correlations between dynamical degrees of freedom $\eta_i(t)$ and the $x_{ij}$ and $c_{ij}$ that characterize disorder, are negligible. This allows factoring of averages of the form $\overline{c_{ij}\,x_{ij}\,\langle \eta_i(t) \rangle}^{c_{ij},x_{ij}}\,\simeq\, \overline{c_{ij}\,x_{ij}}^{c_{ij},x_{ij}}\,\overline{\langle \eta_i(t) \rangle}^{c_{ij},x_{ij}}$. The fraction $m(t+1)$ of failed processes at the next time-step, $t+1$, evaluates to
\begin{eqnarray}
\label{eq:define macro variable}
\nonumber m(t+1) &=& \frac{1}{N}\sum_{j}\eta_j(t+1) \\
&=& \frac{1}{N}\sum_{j}\Theta \left( h_j(t)\,-\vartheta_j\,+\,\xi_j(t)\right)\,.
\end{eqnarray}
In the limit $N\,\to\,\infty$, Eq. (\ref{eq:define macro variable}) can be evaluated by appealing to the law of large numbers as a sum of averages over the random variables $h_j(t)$, $\xi_j(t)$ and $\vartheta_j$. As $h_j(t)$ and $\xi_j(t)$ are Gaussian random variables, their sum is also Gaussian, with mean $J_0\,m(t)$ and variance $1\,+\,J^2\,m(t)$. Thus, first performing the joint average over local fields and noise, we obtain
\begin{equation}
\label{eq:order para before theta avg}
m(t+1)\,=\,\frac{1}{N}\sum_{i}\Phi \left(\frac{J_0\,m(t)\,-\,\vartheta_i}{\sqrt{1\,+\,J^2\,m(t)}} \right)\,.
\end{equation}
Finally, we note that the only $i$ dependence in Eq. (\ref{eq:order para before theta avg}) comes from the $\vartheta_i$. Thus, in the limit $N\,\to\,\infty$, we obtain an average over the $\vartheta$ distribution \footnote{{\it Note that heterogeneity of the averages in equation (\ref{eq:local field average})-(\ref{eq:local field var}), induced through $\vartheta_j$ was left implicit there.}}, which we denote by $\overline{ (\ldots) }^{\vartheta}$, giving
\begin{equation}
\label{eq:macro observable dynamic}
m(t+1)\,=\,\overline{ \Phi \left(\frac{J_0\,m(t)\,-\,\vartheta}{\sqrt{1\,+\,J^2\,m(t)}} \right)}^{\vartheta}\,.
\end{equation}

We have thus obtained a simple closed expression for the evolution of the faction of failed processes in the network.

\subsection{\label{sec: GFA solution} Generating Functional Analysis}
Investigations of the dynamical properties of a system are conducted systematically employing GFA \cite{Dominicis:1978}, which provides tools for the evaluation of correlation and response functions in terms of a characteristic functional of path-probabilities. Performing the average over bond disorder in the sum over dynamical trajectories, one obtains a family of effective single site processes parameterized by $\vartheta$,
\begin{eqnarray}
\label{eq:effective single site process 01}
\nonumber \eta(t+1) &=& \Theta \Big( J_0\,m(t)\, +\, \alpha\,J^2\,\sum_{s<t}G(t,s)\,\eta(s) \\
 & & -\, \vartheta \,+\, \phi(t)\,+\,h(t) \Big)\,.
\end{eqnarray}
These single site processes exhibit memory, via the response function, $G(t,s)$, and are driven by colored Gaussian noise, $\{\phi(t)\}$, self consistently determined via
\begin{eqnarray}
\label{eq:colored noise covar a}
\langle \phi(s)\,\phi(t) \rangle &=& \delta_{[s,t]}\,+\,J^2\,q(s,t)\,,\\
\label{eq:order para after saddle point 01a}
m(t) &=& \overline{ \left\langle \eta(t) \right\rangle }^{\vartheta}\,,\\
q(s,t) &=& \overline{ \left\langle \eta(s)\,\eta(t) \right\rangle }^{\vartheta}\,,\\
\label{eq:order para after saddle point 01b}
G(t,s) &=& \frac{\partial\,m(t) }{\partial\,h(s)}\,,\qquad s\,<\,t\,.
\end{eqnarray}
Here $\langle (\ldots) \rangle$ refers to average over $\{\phi(t)\}$. The external field $h(t)$ is primarily  introduced to define the response function via Eq. (\ref{eq:order para after saddle point 01b}). The derivation and interpretation of these parameters are provided in appendix {\ref{apx: GFA}. The order parameter $m(t)$, equation (\ref{eq:order para after saddle point 01a}), describes the dynamics of the fraction of failed processes.

In the case $\alpha=0$, we recover Eq. (\ref{eq:macro observable dynamic}), which was derived via purely heuristic reasoning in Sec. \ref{sec: Heuristic Solution}. However, for arbitrary $\alpha$, the response function complicates averaging over $\{\phi(t)\}$. In these cases, numerical results for Eqs. (\ref{eq:order para after saddle point 01a})-(\ref{eq:order para after saddle point 01b}) are obtained using Eissfeller-Opper (EO) simulations \cite{Opper:1992}. A key concern in performing the simulations is producing colored noise. Our method uses a Cholesky decomposition \cite{NumRepEd2:2002} of the noise covariance matrix. We construct colored noise as a linear combination of white noises weighted with elements of the Cholesky matrix. 

\subsection{\label{sec: stationary states solution} Equilibrium Statistical Mechanics}
A process network of the form described in Sec. \ref{sec:Model Definition} is guaranteed to achieve a stationary probability distribution for its microscopic states, if we assume $\alpha=1$ and use thermal noise, distributed according to
\begin{equation}
\label{eq: thermal noise}
p(\xi)\,=\,\frac{1}{2}\,\beta\, {\rm sech}^2 \left( \frac{\beta\,\xi}{2} \right)\,,
\end{equation}
instead of Gaussian noise, as in Eq. (\ref{eq:process state}). The parameter $\beta$ is called the inverse temperature. The Gibbs-Boltzmann equilibrium distribution is characterized by the Hamiltonian,
\begin{equation}
\label{eq:gibbs boltz dist Hamiltonian 01}
H(\boldsymbol{\eta})\,=\,-\sum_{i < j}c_{ij}\,\tilde{J}_{ij}\,\eta_{i}\,\eta_{j}\, + \,\sum_{i}\vartheta_{i}\,\eta_{i}\,.
\end{equation}
A solution of this model, with disorder variables $\vartheta_i$, $c_{ij}$ and $x_{ij}$, given by Eqs. (\ref{eq:decompose bond disorder})-(\ref{eq:define x_ij}), is achieved with techniques used to solve the Sherrington-Kirkpatrick (SK) spin-glass \cite{SK:1975} model. In this spirit, the Replica-Symmetric (RS) order parameters are obtained from the extensive free energy as solutions of the following pair of self-consistency equations,
\begin{eqnarray}
\label{eq:m from replica 01}
m &=& \overline{\int{\cal D}z\,\Phi_{\beta}\left(h_{{\rm RS}}\right)}^{\vartheta}\,,\\
\label{eq:q from replica 01}
q &=& \,\overline{\int{\cal D}z\,\Phi_{\beta}^2\left(h_{{\rm RS}}\right)}^{\vartheta}\,,
\end{eqnarray}
where, 
\begin{equation}
\Phi_{\beta}(x) = \frac{1}{2}\left(1\,+\,\tanh\left(\frac{\beta\,x}{2}\right)\right)\,,
\end{equation}
and
\begin{equation}
\label{eq:replicated local field}
h_{{\rm RS}} = -\vartheta\,+\,J_{0}\,m\,+\,J\,\sqrt{q}\,z\,+\,\frac{\beta\,J^{2}}{2}\,(m-q)\,.
\end{equation}
Derivations are provided in appendix \ref{apx: Replica}. The order parameter $m$, Eq. (\ref{eq:m from replica 01}), describes the stationary fraction of failed processes.

The stability of this solution against Replica Symmetry Breaking (RSB) was checked by verifying that the Hessian at the RS saddle point is positive definite \cite{Almeida:1977}. This requires that the so called replicon eigenvalue, given by
\begin{equation}
\label{eq:stable eigenvalues}
\lambda\,=\,1\,-\,(\beta\,J)^2\int{\cal D}z\,\left(\Phi_{\beta}(h_{{\rm RS}})\,-\,\Phi_{\beta}(h_{{\rm RS}})^2\right)^2\,
\end{equation}
is positive. We found that for networks with parameter settings as investigated in section \ref{sec: Results} below, i.e., a-priori homogeneous failure probability, $p=0.01$, RSB occurs only for $J \gtrsim 4.630$. Hence, the regime of interest with small frustration, say $J<0.5$, is well within the stable domain.
 
Comparison between the equilibrium $m$, Eqs. (\ref{eq:m from replica 01})-(\ref{eq:replicated local field}), and the long-term stationary behavior of $m(t)$, Eq. (\ref{eq:order para after saddle point 01a}), with $\alpha=1$ is possible, once we scale the thermal noise appropriately to match properties of the Gaussian noise. In principle, such a matching may be accomplished in several different manners, each of which leads to similar results. Here, we prescribe that the thermal noise should have the same unit variance as the Gaussian noise used in the microscopic dynamics. This leads to $\beta\,=\,\pi/\sqrt{3}$.

\section{\label{sec: Results} Results}
In this section we explore the consequences of theory developed above, with particular emphasis on phase diagrams of resulting stationary states. 

For simplicity, we assume homogeneous $\vartheta$, which allows us to drop the corresponding $\vartheta$-averages. We justify this simplification by noting that any other non-trivial $\vartheta$-distribution, for example Gaussian with mean $\overline{\vartheta}$ and variance $\sigma_{\vartheta}^2$, would not alter the qualitative behavior of the observables. We illustrate this for the fully asymmetric network, $\alpha=0$. To evaluate Eq. (\ref{eq:define macro variable}) we exploit that the sum of Gaussian terms, $h_j(t)\,-\,\vartheta_j\,+\,\xi_j(t)$, is itself a Gaussian with mean $J_0\,m(t)\,-\,\overline{\vartheta}$ and variance $1\,+\,J^2\,m(t)\,+\,\sigma_{\vartheta}^2$. This gives
\begin{equation}
\label{eq: m of t+1 with gaussian vartheta}
m(t+1) \,=\, \Phi \left(\frac{J_0\,m(t)\,-\,\overline{\vartheta}}{\sqrt{1\,+\,J^2\,m(t)\,+\,\sigma_{\vartheta}^2}} \right)\,.
\end{equation}
Up to a shift in the variance of the noise, this equation describes the evolution of a system of asymmetrically coupled processes with uniform $\vartheta_i\,=\,\overline{\vartheta}$.

For stationary states for the fully symmetric network, $\alpha=1$, obtained from equilibrium statistical mechanics, a similar structural shift can be shown to apply.

\subsection{Fully Asymmetric Network}
Assuming long-term stationary behavior for the observable $m(t)$, Eq. (\ref{eq:macro observable dynamic}), we drop the time index. The curve in the first panel of Fig. \ref{fig:asym} depicts $m$ as a function of $J_0$. We note the coexistence of a low-$m$ operational and a high-$m$ non-operational phase for $J_0$ values bounded by the lower and upper critical values, $J_0^c\,\approx\,3.971$ and $J_0^c\,\approx\,14.814$, respectively. Within this interval there is an unstable branch of solutions, with intermediate $m$-values, represented by the back-bending part of the curve.

The behavior is parameterized by the unconditional failure probability $p$. This dependence is made explicit in the phase diagram, shown in the second panel of Fig. \ref{fig:asym}. The upper curve marks the $J_0^c$-boundary of the low-$m$ operational phase, while the lower curve represents the $J_0^c$-bound of the high-$m$ non-operational phase. A transition between the two phases is discontinuous for $p\,<\,p_c \approx 0.102$ and becomes continuous exactly at $p_c$. Beyond $p_c$ the phases loose their separate identity, much as in a liquid-gas system. In the region of small $J$, which is of interest for OR, say $J\,<\,0.5$, the behavior shown in Fig. \ref{fig:asym} is fairly insensitive to variation in $J$.
\begin{figure}
\includegraphics[scale=0.7]{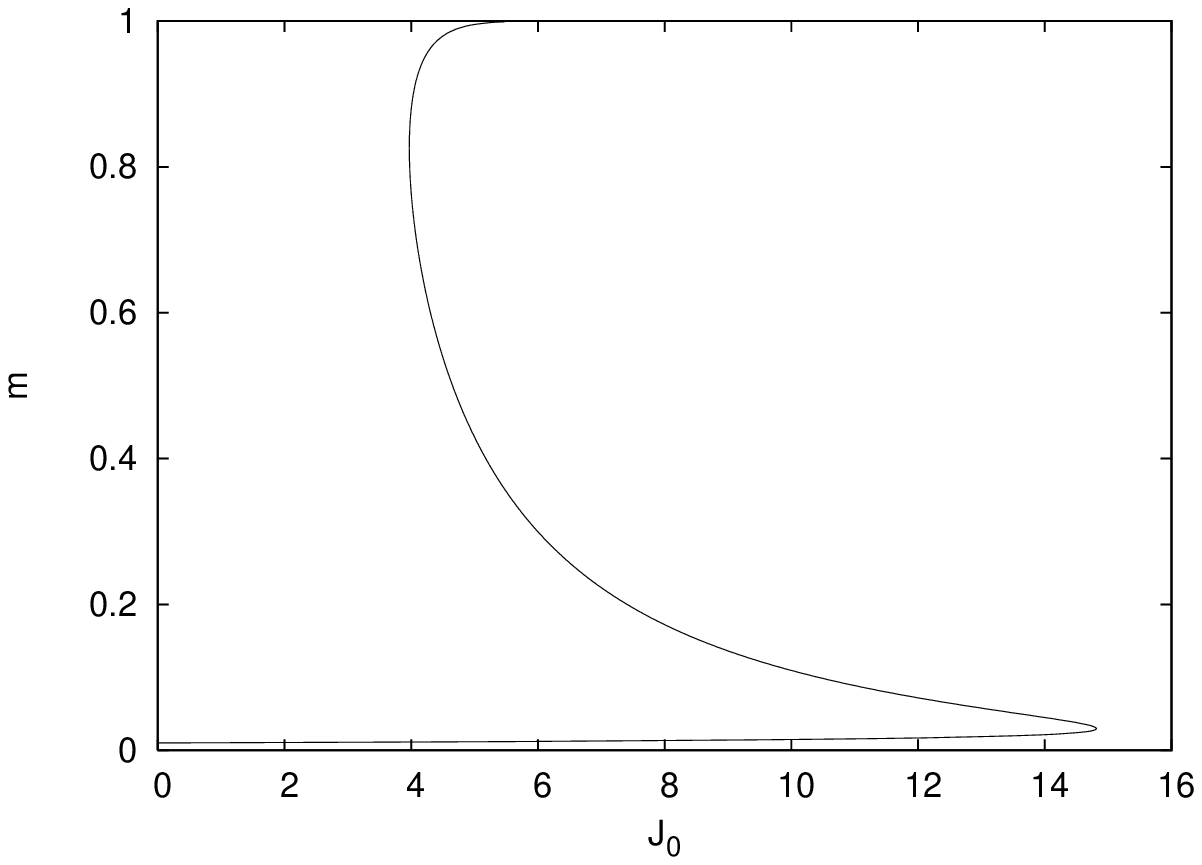}
\includegraphics[scale=0.7]{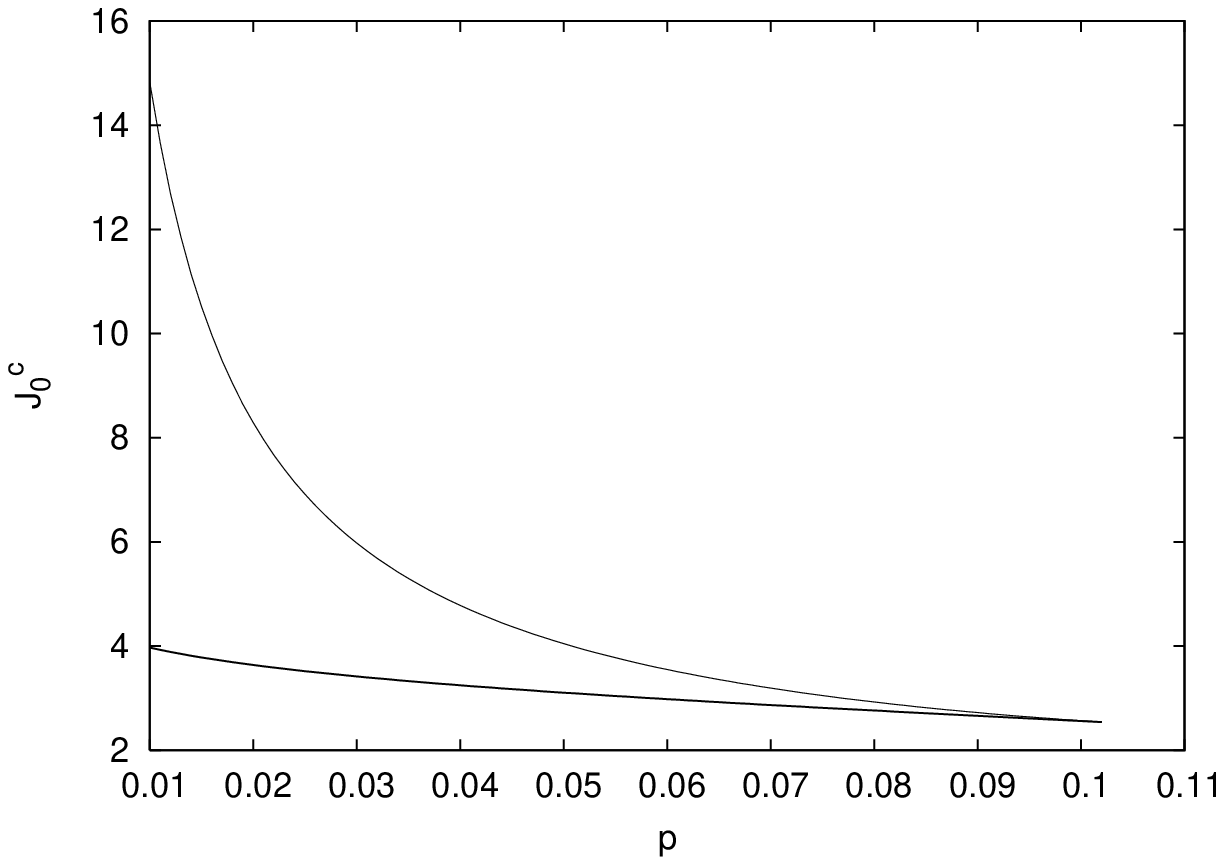}
\caption{\label{fig:asym} First panel: Stationary observable $m$ as a function of $J_0$. The homogeneous unconditional failure probability is $p=0.01$. Second panel: Phase diagram showing critical values of $J_0$, as a function of $p$, where the operational (upper curve) and non-operational (lower curve) phases become unstable. In producing both diagrams, we set $J=0.2$.}
\end{figure}

\subsection{\label{subsec: partial sym network}Partially Symmetric Network}
Investigations of partially symmetric networks were conducted via: (i) evaluating the first three time-steps of the effective single site dynamics exactly, starting with random initial conditions, given by $m(0)$, and (ii) conducting EO simulations of the dynamics to higher time-steps.

The results presented in Fig. \ref{fig:gfa}, were produced for an intermediate value of the symmetry parameter $\alpha=0.5$.

The exact evaluation of the first three time-steps is depicted in the first panel. The curves represent the fraction of failed processes at each time as a function of $J_0$. We observe that these curves intersect the line $m(0)=0.3$ at exactly the same point, $J_0^T \approx 6.001$, which marks a change in behavior.  For $J_0 < J_0^T$, a convergence to the low-$m$ operational phase is realized, while for $J_0 > J_0^T$, we observe convergence to the high-$m$ non-operational phase. Except in the immediate vicinity of $J_0^T$, the convergence is rapid and achieved within the first three time-steps.

To investigate the behavior beyond the first three time-steps, we used EO simulations to propagate Eqs. (\ref{eq:effective single site process 01})-(\ref{eq:order para after saddle point 01b}) up to $t=30$. This time horizon is sufficient to achieve stationarity, except in an infinitesimal neighborhood of the critical $J_0^c$ values. In the second panel of Fig. \ref{fig:gfa} we provide the stationary $m$ value as a function of $J_0$. The coexistence of the low-$m$ operational phase and high-$m$ non-operational phase is observed for intermediate $J_0$, in complete analogy to the fully asymmetric case. These $J_0$ values are bounded between the critical values, $J_0^c \approx 3.966$ and $J_0^c \approx 14.513$. Here, the unstable branch, represented by the back-bending curve, is computed by locating the $J_0^T$ values that separate the regions of convergence towards either the operational or non-operational phase. It is noteworthy that we obtain an almost identical unstable branch if we use the exact forms of the first three time-steps.
\begin{figure}
\includegraphics[scale=0.7]{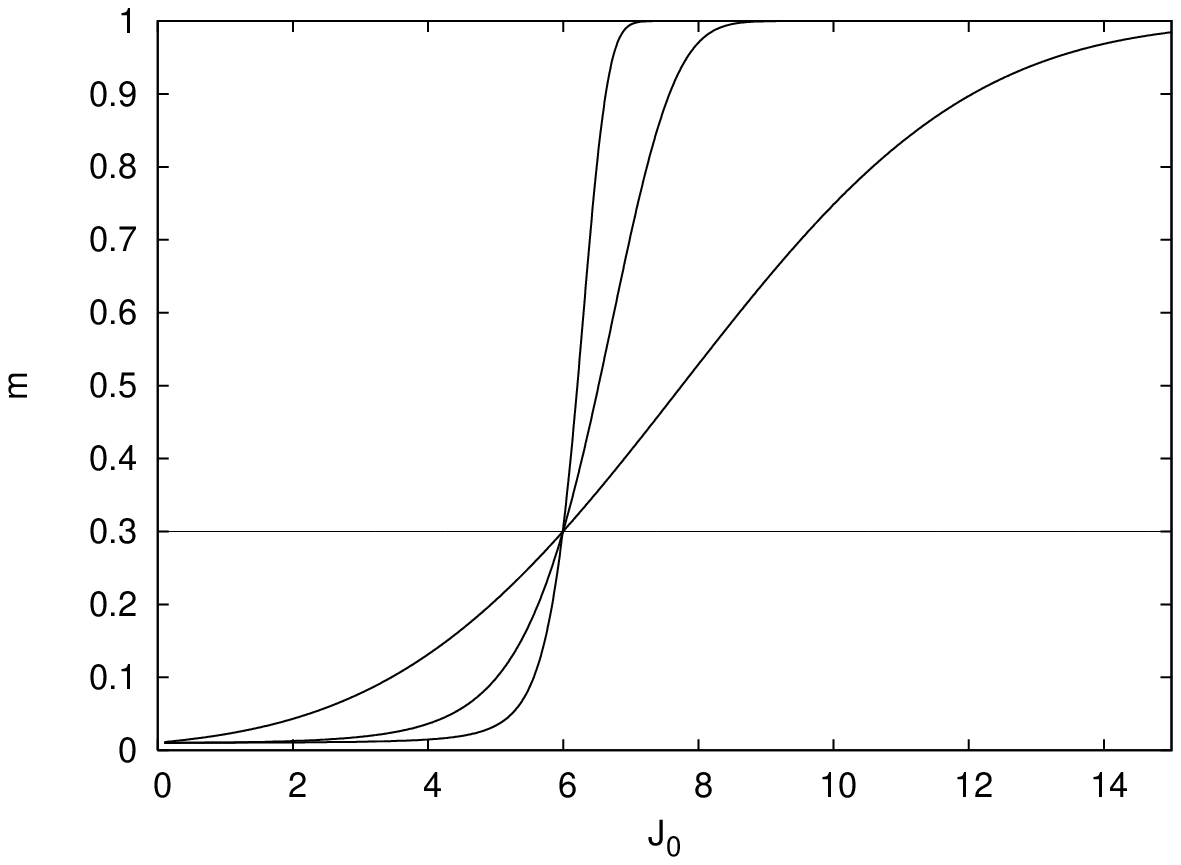}
\includegraphics[scale=0.7]{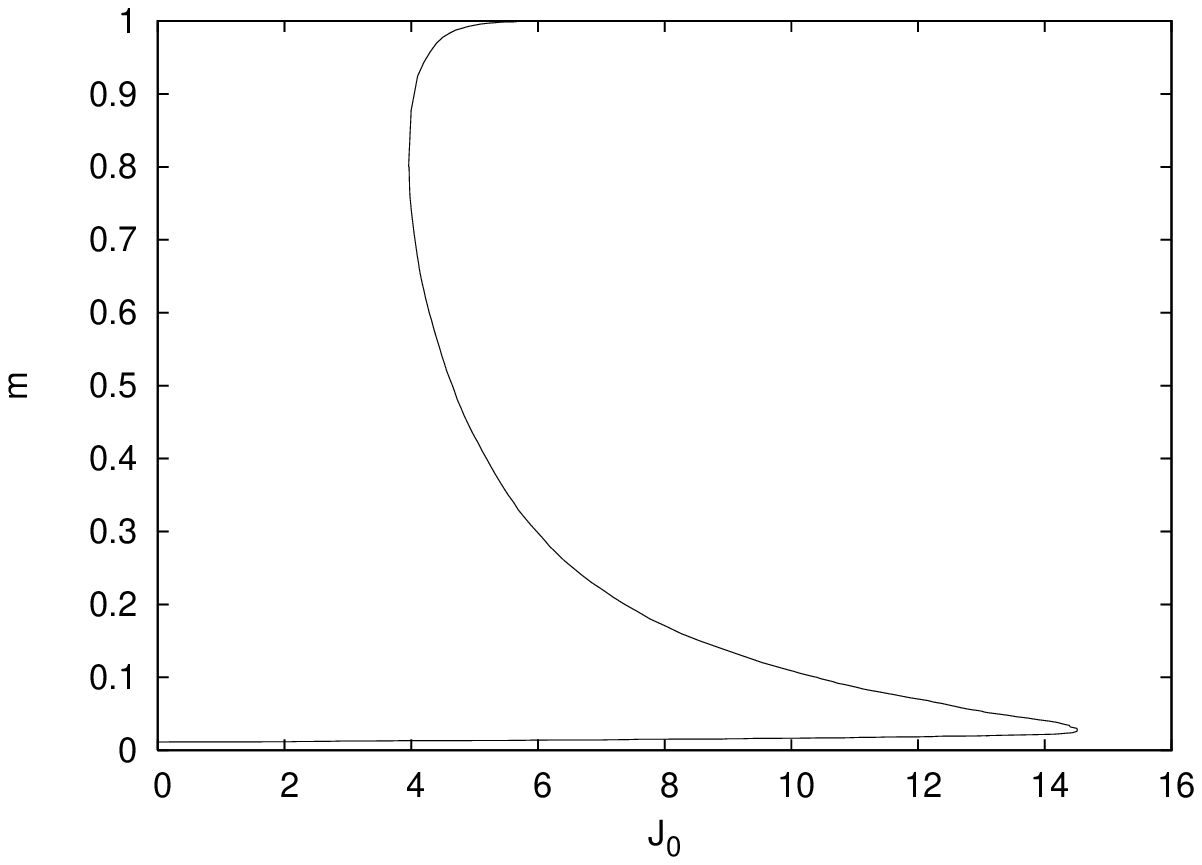}
\caption{\label{fig:gfa} First panel: First three time steps of $m(t)$ for $\alpha=0.5$ and initial condition $m(0)=0.3$, as a function of $J_0$. The curves $m(1)$, $m(2)$ and $m(3)$ are given in order of increasing steepness about their intersection with $m(0)$. Second panel: Stationary value of $m$ as a function of $J_0$ from EO simulations, with $\alpha=0.5$ and $p=0.01$. }
\end{figure}

\subsection{\label{subsec:stationary results}Fully Symmetric Network}
Investigations of fully symmetric networks, $\alpha=1$, were conducted via: (i) appealing to equilibrium statistical mechanics, Eqs. (\ref{eq: thermal noise}) - (\ref{eq:replicated local field}), and (ii)  EO simulations. 

To compare the two results, we scaled the thermal noise to have unit variance, resulting in $\beta = \pi/\sqrt{3}$.

In the first panel of Fig. \ref{fig:replica}, we plot values of $m$ as a function of $J_0$. The dotted line was produced using techniques from equilibrium statistical mechanics. The solid line was produced from EO simulations for $\alpha=1$. In both cases, the coexistence of a low-$m$ operational and a high-$m$ non-operational phase, for intermediate values of $J_0$, is observed. The coexistence regions are given by the intervals $[4.113,\,20.478]$ and $[3.966,\,14.512]$ for the networks with thermal and Gaussian noise, respectively. Whereas the critical values for the non-operational phase are almost identical, a significant discrepancy for the upper critical value is observed. This discrepancy must be attributed to differences in the nature of the noises. 

In the second panel of Fig. \ref{fig:replica}, we produce the phase diagram for the system with thermal noise. We observe a similar qualitative behavior to the fully asymmetric case, $\alpha=0$, given in Fig. \ref{fig:asym}. The critical point $p_c \approx 0.110$ is very close to that from the asymmetric case, $p_c \approx 0.102$.
\begin{figure}
\includegraphics[scale=0.7]{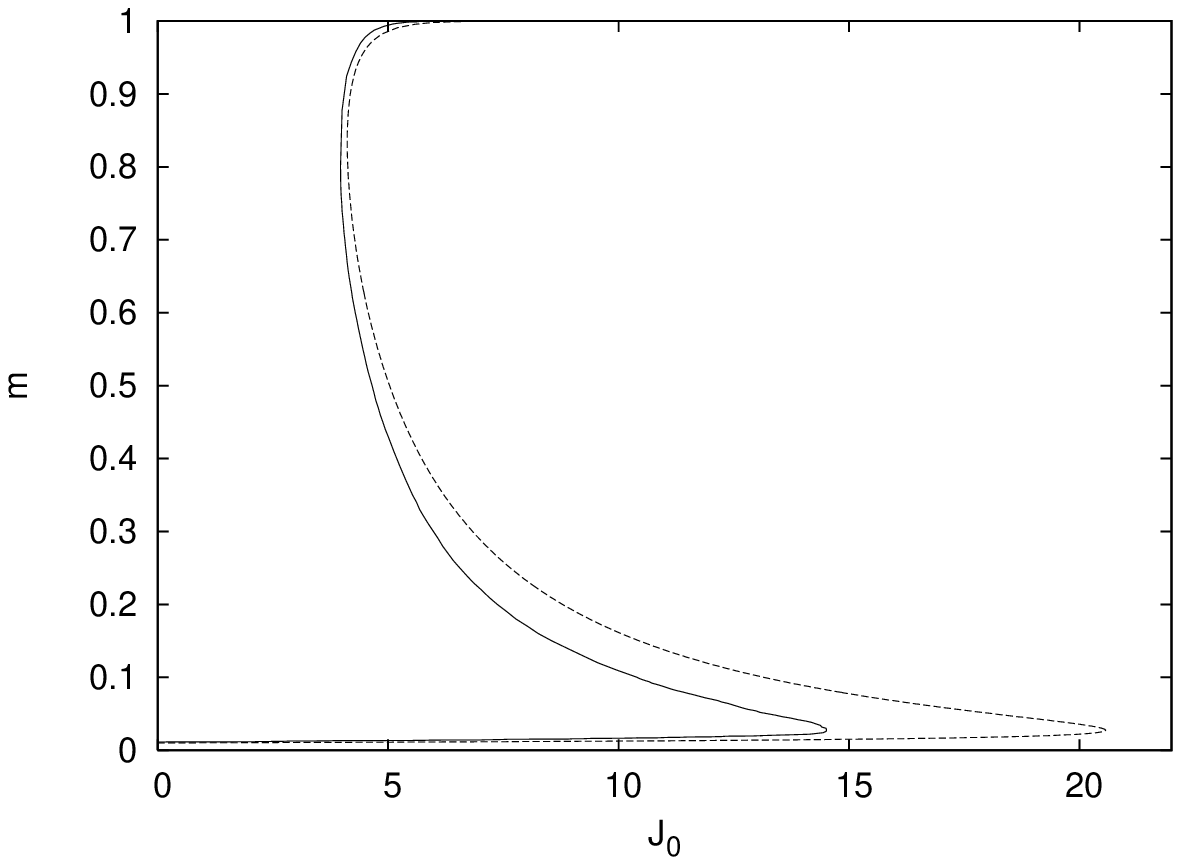}
\includegraphics[scale=0.7]{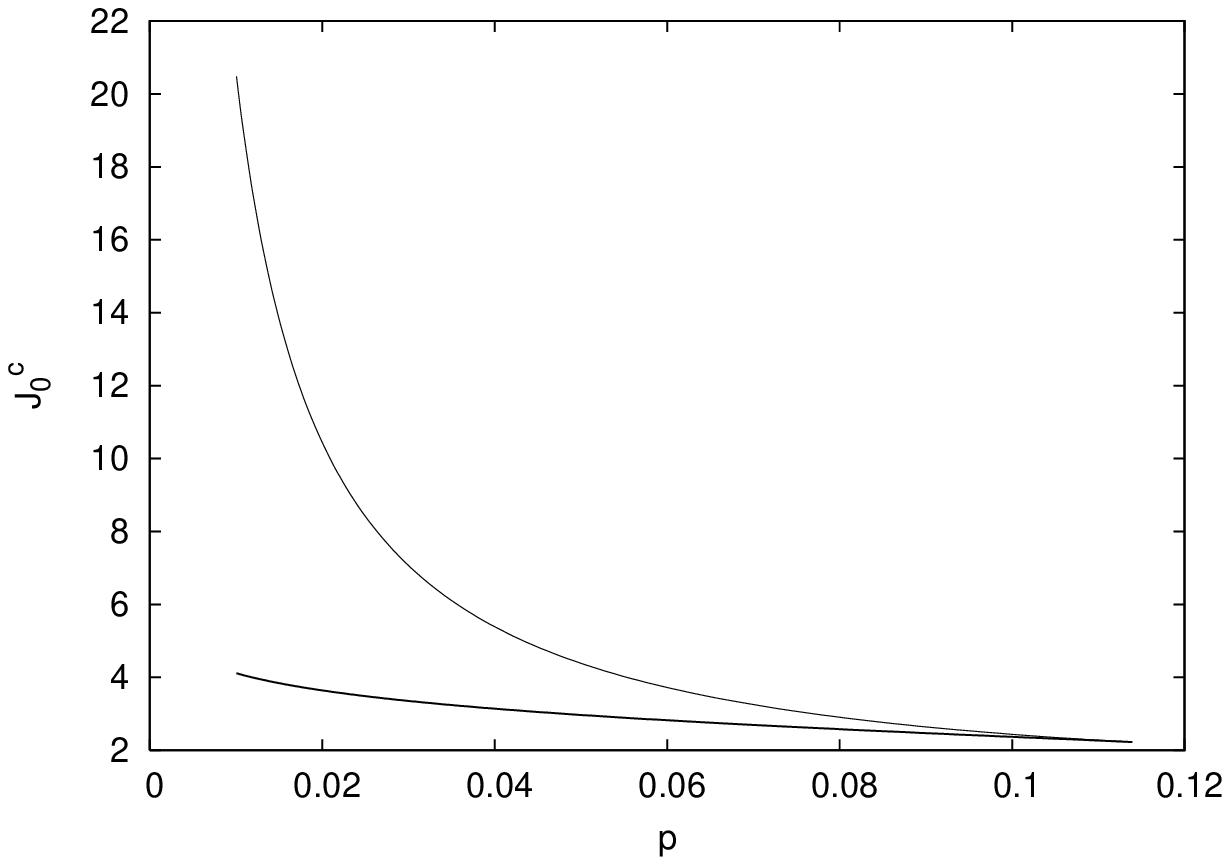}
\caption{\label{fig:replica} First panel: Comparisons between stationary values of $m$ for $\alpha=1.0$ for networks with Gaussian noise (solid line) to networks with thermal noise (dotted line), as functions of $J_0$. The inverse temperature was $\beta=\pi/\sqrt{3}$, and the a-priori homogeneous failure probability is $p=0.01$. Second panel: Phase diagram of the network with thermal noise. In producing both diagrams, we set $J=0.2$.}
\end{figure}

\subsection{Loss Distribution and Capital Requirement}
Finally, in Fig. \ref{fig:lossdist} we show a loss distribution for a network of $50$ processes. The stationary macroscopic variable $m$ represents the fraction of failed processes. Alternatively, $m$ represents the probability that a process breaks down within an interacting environment. That is, effects of functional correlations with other processes in the network have been accounted. We proceed by drawing from a binomial distribution, for each process $i$ the number $k_i$ of days in a year the process fails. The daily failure probability is given by $m$. Failures on different days are taken to be independent. Subsequently binning the sum of $k_i$ realizations from the log-normal loss severity distribution for each process, we obtain the loss distribution for the network.

In the thermodynamic limit one would expect the loss distribution to be a Gaussian by the central limit theorem. However, for $N=50$ the distribution is still heavily skewed to the right with a ``fat" tail and rare realizations of extremely large losses.
\begin{figure}
\includegraphics[scale=0.7]{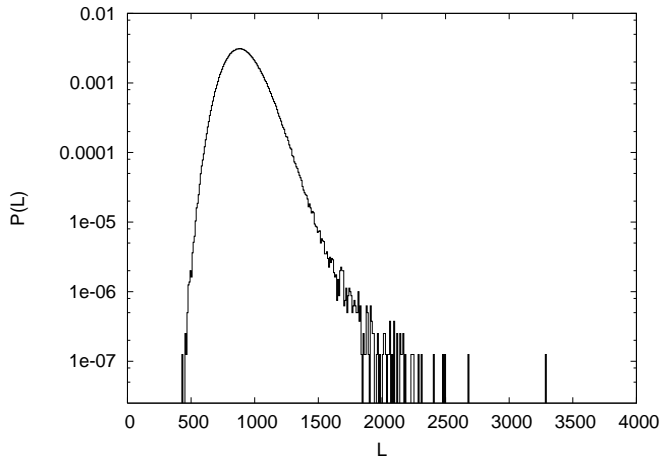}
\caption{\label{fig:lossdist} Loss distribution for a network of $50$ processes constructed using the LDA. A binomial distribution, with mean $m=0.01$, was used for the loss frequency distribution. A log-normal distribution, with mean $\mu = 5$ and variance $\sigma^2 = 75$, was used for the loss severity distribution.}
\end{figure}

A qualitative study was also conducted to determine the effects of functional correlations on the VaR. Recall that under LDA, VaR is used to determine the capital to be allocated for ORs. We consider a fully asymmetric network with homogeneous unconditional and conditional failure probabilities. First, for a non-interacting environment, the loss frequency distribution is binomial with daily failure probability, $m = \phi(\vartheta)$. Using the construction as described afore, the loss distribution was evaluated and ${\rm VaR}_0$ calculated for a confidence level $q = 99.99\%$. Next, for the interacting environment, the loss frequency distribution is once more binomial. However the daily failure probability, $m$ is now given by the stationary operational solution to Eq (\ref{eq:macro observable dynamic}), as a function of $J_0$. The loss distributions are evaluated and the VaR determined for the same confidence level. In Fig. (\ref{fig:VaR_comp}) we plot the ratio VaR to ${\rm VaR}_0$ as a function of $J_0$. A significant increase in the VaR is clearly discernible. This is mainly driven by interactions induced  through increasing $J_0$, which enlarges the daily failure probability, $m$. As Fig. (\ref{fig:VaR_comp}) illustrates, for a company with a high degree of interdependence between processes, ${\rm VaR}_0$ does not capture the true extent of risk. 
\begin{figure}
\includegraphics[scale=0.7]{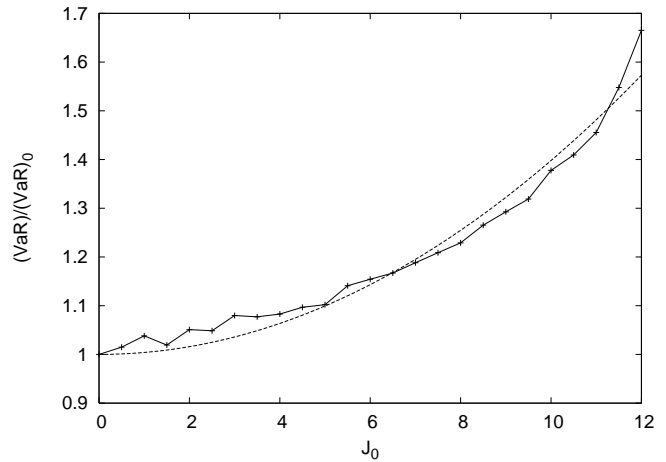}
\caption{\label{fig:VaR_comp} Ratio of VaR to ${\rm VaR}_0$, as a function of $J_0$. We consider a network of $50$ processes with $J = 0.2$, over a risk horizon of $365$ days. The data is fitted against a quadratic curve, represents by the dashed line.}
\end{figure}

\section{\label{sec: Conclusion} Conclusion}
In this paper we investigated networks of interacting and functionally correlated processes as models of OR in large organizations. In the limit of sparse connectivity, we studied the dynamical evolution and stationary states of these networks using analytical methods and numerical simulations. A heuristic approach was used to exactly solve the dynamics of fully asymmetric networks. GFA and EO simulations were used to study the dynamics of partially and fully symmetric networks. Finally, stationary states for fully symmetric networks were analyzed using techniques from equilibrium statistical mechanics.

The initial variant of our model was formulated on a sparsely connected random graph, for which the average connectivity per process $c$, satisfies, $c/N \to 0$ in the thermodynamic limit. However, it should be noted that the assumption needed to carry through the GFA as described in this paper is $c \gg 1$ rather than sparseness. Hence, the analysis applies to networks with non-sparse connectivity with $c = {\cal O}(N)$ as well.

From all investigated cases, we found that there exists a range of interactions, characterized by $J_0$, for which an operational phase coexists with a non-operational phase. This raises the possibility of spontaneous catastrophic breakdown in such process networks. We demonstrated this feature is robust and invariant under a broad range of changes in microscopic details.

Phase diagrams were produced for fully asymmetric networks using the exact dynamical theory and for fully symmetric networks using the results from equilibrium statistical mechanics. Similar behavior is observed in both cases. For sufficiently small values of a-priori failure probability $p$, the operational and non-operational phases are distinct; a discontinuous transition between phases can be induced, for instance, by changing $J_0$. There is a critical failure probability $p_c$, where the transition is continuous. We also note that the two critical values, $p_c \approx 0.102$ for the fully asymmetric case, and $p_c \approx 0.110$ for the fully symmetric case are very close.

Given parameter values $p$, $J_0$ and $J$ for which {\em stable\/} operational and non-operational phases coexist, one can further qualify these phases as either {\em absolutely stable\/} or {\em meta-stable\/}. These notions are well known for thermodynamic equilibrium systems. Therein, the absolutely stable phase is the one with the lower value of the free energy. Parameter values at which absolutely stable phases become meta-stable (or vice-versa) are the locations of proper first-order equilibrium phase transitions.

Of the systems studied in this paper, however, only the fully symmetric network with thermal noise can be characterized as a thermodynamic equilibrium system. Nevertheless, the notion of absolute stability and meta-stability of coexisting dynamic stationary states can be carried over to the other systems in the following manner. One denotes by $\tau_N$ the fraction of the total time a system of size $N$ spends in one of the coexisting stationary states, either operational or non-operational. A state is characterized as absolutely stable if $\lim_{N\to \infty}\tau_N =1$ , and as meta-stable, if $\lim_{N\to \infty}\tau_N = 0$. A meta-stable phase can nevertheless be dynamically stable in the sense that, once in such a phase, a system will typically stay in that phase for a very long time $t_N$ which diverges as $N\to \infty$.

These concepts have a significant implication for OR, that play out for large (but finite) process networks that exhibit dynamically and absolutely stable operational phases, coexisting with a meta-stable non-operational phases. Small changes of the parameters characterizing the network, e.g. a slight increase in the average mutual dependency among processes --- described in our model by a slight increase in $J_0$ --- could entail that the operational phase becomes {\em meta-stable\/}. In this case a  catastrophic breakdown is bound to occur, even under normal operating conditions. A small increase in $J_0$ would result in a correspondingly small change in the statistical properties of the network. As such, there would be no visible precursors for the transition. In other, more colorful words, a small change in the design of a process network could amount to inadvertently entering a gate carrying the inscription ``{\it Lasciate ogni speranza, voi ch'entrate}" \cite{Dante}.

Of particular relevance in this context are a number of current trends such as the advance of globalization, increasing reliance on information technology, or the growing modularization of business  and production processes, including out-sourcing. Within our model these trends roughly correspond to a trend of increasing 
$J_0$; for a stable operational phase this is equivalent to a trend of pushing that phase towards meta-stability, hence eventually towards guaranteed catastrophic breakdown.

A recent incident involving the retail bank HSBC outsourcing business processing highlights this issue \cite{BBC}. The outsourced data included bank account passwords, along with other sensitive information to an office in Bangalore, India. A widespread fraud, that took roots in 2002 and affected one thousand of the banks' clients was uncovered earlier this year. While this did not critically affect the bank, its vulnerability has increased. The fact that sensitive information was out-sourced to a single office shifted the banks' stable operational phase further towards the meta-stability.
 
It is imperative that banks and other organizations assess their stability, as described above, and check for the possible coexistence of operational and non-operational phases. As there are no detectable precursors for transitions between them, the assessment must be performed by stress tests, wherein artificial strains and fluctuations are introduced into the system and the effects are measured. The present model readily lends itself to tests of this type.

In our investigation, we only considered Poisson random graphs with large average connectivity $c$. As a consequence, there is little heterogeneity in the local environment about each process. However, it is known that realistic process networks exhibit a richer topology, which includes the presence of ``hubs". In such situations, the connectivity scales according to a power-law and the analytical treatment would follow lines of reasoning used in small-world networks \cite{Dorogovtsev:2003}. A similar qualitative behavior with coexistence of operational and non-operational phases is nevertheless expected.   

\appendix
\section{\label{apx: GFA} Generating Functional Analysis}
In this appendix we use Generating Functional Analysis (GFA) \cite{Dominicis:1978} to formally solve the model dynamics and  systematically justify the heuristic solution in Sec. \ref{sec: Heuristic Solution}. We begin by introducing the generating functional over source fields, $\boldsymbol{\psi}$,
\begin{equation}
\label{eq:generating function}
Z[\boldsymbol{\psi}] \,=\, \left\langle\, \exp\left(-{\rm i}\,\sum_{i=1}^N\,\sum_{t=0}^T\,\psi_{i}(t)\,\eta_{i}(t)\right)\,\right\rangle\,.
\end{equation}
The angled brackets refer to the average over all ``paths", which are trajectories of microscopic states. Explicitly,

\begin{eqnarray}
\label{eq:generating function with explicity path probability}
Z[\boldsymbol{\psi}] = \sum_{\{\boldsymbol{\eta}(t)\}}P[\{\boldsymbol{\eta}(t)\}]
\exp\left(-{\rm i}\sum_{t=0}^{T}\sum_{i=1}^{N}\,\psi_{i}(t)\eta_{i}(t)\right)
\end{eqnarray}
where $P[\{\boldsymbol{\eta}(t)\}]\,=\,P[\,\boldsymbol{\eta}(0),\ldots,\boldsymbol{\eta}(T)\,]$ denotes the probability of all paths over the risk horizon, $T$. The generating functional can be used to compute expectation values and correlation functions as

\begin{eqnarray}
\label{eq:generating function - compute expectation and correlation}
\langle \eta_i(t) \rangle &=& {\rm i}\left.\frac{\partial Z[\boldsymbol{\psi}]}{\partial \psi_i(t)}\right\vert_{\boldsymbol{\psi} \equiv 0}\,,\\
\langle \eta_j(s)\,\eta_i(t) \rangle &=& {\rm i}^2\left.\frac{\partial Z[\boldsymbol{\psi}]}{\partial \psi_j(s)\,\partial \psi_i(t)}\right\vert_{\boldsymbol{\psi} \equiv 0}\,.
\end{eqnarray}

We are interested in evaluating the generating functional for a typical realization of disorder. This is achieved by averaging Eq. (\ref{eq:generating function with explicity path probability}) over $c_{ij}$ and $x_{ij}$. To proceed, we exploit that the path probability measure has a Markov structure.
\begin{equation}
\label{eq:path prob markov defn}
P[\{\boldsymbol{\eta}(t)\}] = P(\,\boldsymbol{\eta}(0)\,)\, \prod_{t=0}^{T-1} P(\,\boldsymbol{\eta}(t+1)\, |\, \boldsymbol{\eta}(t)\,)\,,
\end{equation} 
with transition probabilities
\begin{equation}
\label{eq:transition probability}
P(\,\boldsymbol{\eta}(t+1)\, |\, \boldsymbol{\eta}(t)\,)\,=\,\prod_{i=1}^{N}\,\int\,{\cal D}\,\xi_{i}(t)\,\delta_{[\eta_{i}(t+1)\,,\,f_{i}(t)]}\,.
\end{equation}
The $\xi_i(t)$ are unit mean and zero variance Gaussian random variables. Furthermore,
\begin{equation}
\label{eq:f i t}
f_{i}(t)\,=\,\Theta\,\left(\, \sum_{j}c_{ij}\,\tilde{J}_{ij}\,\eta_{j}(t)\,+\,h_i(t)\,-\,\vartheta_{i}\,+\,\xi_{i}(t) \,\right)\,.
\end{equation}

\subsection{Average Over Fast Noise}
We conduct the $\xi_i(t)$ integral and disorder average by first extracting these contributions $u_i(t)\,=\,\sum_{j}c_{ij}\,\tilde{J}_{ij}\,\eta_{j}(t)\,+\,\xi_{i}(t)$ from the Heaviside function in $f_i(t)$. This is facilitated utilizing a Dirac $\delta$-function and its normalization property to give us
\begin{widetext}
\begin{eqnarray}
\label{eq:GF after fast noise avg}
\nonumber Z[\boldsymbol{\psi}] &=& \sum_{\boldsymbol{\eta}(0)}\,\ldots\,\sum_{\boldsymbol{\eta}(T)}\,P(\,\boldsymbol{\eta}(0)\,)\,\prod_{t=0}^{T-1}\prod_{i=1}^{N}\bigg[ \int \frac{{\rm d}\widehat{u}_i(t)\,{\rm d}u_i(t)}{2\pi} \exp\Bigg( -{\rm i}\,\widehat{u}_i(t)\Bigg(\, u_i(t)  -\,\sum_j c_{ij}\,\tilde{J}_{ij}\eta_j(t)\,\Bigg)\\
& &  -\, \frac{\widehat{u}_i(t)^2}{2} \Bigg) \delta_{\left[\eta_{i}(t+1) \,,\,f_i(t)\right]} \bigg]\,\exp\left(-{\rm i}\sum_{t=0}^{T}\sum_{i=1}^{N}\,\psi_{i}(t)\eta_{i}(t)\right)\,.
\end{eqnarray}
\end{widetext}
In equation (\ref{eq:GF after fast noise avg}), we redefined $f_i(t)$ to denote,
\begin{equation}
\label{eq:f i t with u i t}
f_{i}(t)\,=\,\Theta\,\left(\, u_i(t)\,+\,h_i(t)\,-\,\vartheta_{i}\,\right)\,.
\end{equation}
The ``hatted" conjugate terms, $\widehat{u}_i(t)$, are a consequence of a Fourier representation of the $\delta$-function.
\subsection{Disorder Average}
\label{sec:GFA disorder avg}
The disorder average, which factorize in pairs, affects $c_{ij}$ and $x_{ij}$. We localize the terms involved in the following definition
\begin{eqnarray}
\nonumber \overline{D}^{c_{ij},x_{ij}} &=& \prod_{i < j} \overline{\exp \Bigg( \sum_{t=0}^{T-1}\,c_{ij}\Bigg( {\rm i}\, \widehat{u}_i(t)\,J_{ij}\,\eta_j(t)}\\
\label{eq_disorder_in_Z}
& & \overline{ +\,{\rm i}\,\widehat{u}_j(t)\,J_{ji}\,\eta_i(t)\Bigg)\Bigg)}^{c_{ij},x_{ij}}\,.
\end{eqnarray}

We first perform the $c_{ij}$ average, and proceed by taking the Taylor expansion of the exponential in the limit $c \gg 1$. Next, taking the $x_{ij}$ average, re-exponentiating and keeping the dominant terms, we obtain
\begin{eqnarray}
\nonumber \overline{D}^{c_{ij},x_{ij}} &=& \exp\Bigg(N\Bigg[J_0\,\sum_{t=0}^{T-1}k(t)\,m(t)\\
\nonumber  & &+\,\frac{J^2}{2}\,\sum_{s,t=0}^{T-1} \Bigg( Q(s,t)\,q(s,t)\\
\label{eq:after disorder avg}
& &+\,\alpha\,G(s,t)\,G(t,s)\Bigg)\Bigg]\Bigg)\,,
\end{eqnarray}
which depends on a set of macroscopic variables defined as
\begin{eqnarray}
\nonumber m(t) &=& \frac{1}{N}\sum_{i=1}^{N}\eta_i(t)\,, \\
\nonumber k(t) &=& \frac{1}{N}\sum_{i=1}^{N}{\rm i}\,\widehat{u}_i(t)\,,\\
\nonumber q(s,t) &=& \frac{1}{N}\sum_{i=1}^{N}\eta_i(s)\,\eta_i(t)\,,\\
\nonumber Q(s,t) &=& \frac{1}{N}\sum_{i=1}^{N}{\rm i}\,\widehat{u}_i(s)\,{\rm i}\,\widehat{u}_i(t)\,,\\
\nonumber G(t,s) &=& \frac{1}{N}\sum_{i=1}^{N}{\rm i}\,\widehat{u}_i(s)\,\eta_i(t)\,. 
\end{eqnarray}

To achieve site factorization in the evaluation of $\overline{Z[\boldsymbol{\psi}]}^{c_{ij},x_{ij}}$, one proceeds as usual by enforcing the above definitions of the macroscopic variables via Dirac $\delta$-function identities and their Fourier representations. This subsequently generates a set of conjugate variables to the above set of macroscopic variables. The averaged generating functional is expressed in the following compact form
\begin{equation}
\label{eq: avg generating function in compact form}
\overline{Z[\boldsymbol{\psi}]}^{c_{ij},x_{ij}}\,=\,\int{\cal D}\{\ldots\}\exp\left\{N\left[\Xi_1\,+\Xi_2\,+\,\Xi_3\right]\right\}\,.
\end{equation}
Here, ${\cal D}\{\ldots\}$ represents taking the integral over all the macroscopic order parameters, and their conjugates. The functions $\Xi_1$, $\Xi_2$ and $\Xi_3$, appearing in the exponential of Eq. (\ref{eq: avg generating function in compact form}), are defined as
\begin{eqnarray}
\nonumber \Xi_{1} &=& J_0\sum_{t=0}^{T-1}k(t)\,m(t)\, +\, \frac{J^2}{2}\sum_{s,t=0}^{T-1}\Bigg(Q(s,t)\,q(s,t)\, \\
\label{eq: Xi 1}
& &+\, \alpha\,G(t,s)\,G(s,t)\,,\\
\nonumber \Xi_{2} &=& {\rm i}\sum_{t=0}^{T-1}\left( m(t)\,\widehat{m}(t)\,+\,k(t)\,\widehat{k}(t)\right)\\
\nonumber & & \,+\, {\rm i}\sum_{s,t=0}^{T-1}\Bigg( q(s,t)\,\widehat{q}(s,t)\,+\,Q(s,t)\,\widehat{Q}(s,t)\\ 
\label{eq: Xi 2}
& &+\,G(t,s)\,\widehat{G}(t,s)\Bigg)\,, \\
\nonumber \Xi_3 &=& \frac{1}{N}\sum_{i=1}^{N}\log\,\Bigg\{ \sum_{\{\eta(t)\}}{\rm P}(\eta(0))\\
\nonumber & & \int\,\prod_{t=0}^{T-1}\bigg[\frac{{\rm d}\widehat{u}(t)\,{\rm d}u(t)}{2\pi}\,\delta_{[\eta(t+1),f_i(t)]}\bigg]\\  
\label{eq: Xi 3}
& &  \exp\left(-{\cal S}\,-\,{\rm i}\sum_{t=0}^{T}\psi_i(t)\,\,\eta(t)\right) \Bigg\}\,.
\end{eqnarray}
Here ${\cal S}$ denotes the dynamic action,
\begin{eqnarray}
\nonumber {\cal S} &=& \sum_{t=0}^{T-1}\Bigg( -{\rm i}\widehat{u}_i(t)\,u_i(t) \,-\, \widehat{u}_i(t)^2/2\, -\, {\rm i}\widehat{m}(t)\,\eta_i(t)\\
\nonumber & &-\, {\rm i}\widehat{k}(t)\,{\rm i}\widehat{u}_i(t)\Bigg) \,+\, \sum_{s,t=0}^{T-1} \Bigg( - {\rm i}\widehat{q}(s,t)\,\eta_i(t)\,\eta_i(s) \\
\nonumber & &- {\rm i}\hat{Q}(s,t)\,{\rm i}\widehat{u}_i(t)\,{\rm i}\widehat{u}_i(s)\,-\,  {\rm i}\widehat{G}(t,s)\,{\rm i}\widehat{u}_i(s)\,\eta_i(t) \Bigg)\,.
\end{eqnarray}

Thus, we have transformed the generating functional into an integral with leading order in the exponential of $N$, which can b evaluated using the saddle point technique. The contribution $\Xi_3$, Eq. (\ref{eq: Xi 3}) describes an ensemble of independent dynamical processes. 

\subsection{Saddle Point Equations}
At the saddle point, the macroscopic observables of interest resolve to
\begin{eqnarray}
\nonumber m(t) &=& \frac{1}{N}\,\sum_{i=1}^{N}\langle\,\eta(t)\,\rangle_{(i)}\,,\\
\label{eq:resolved observables}
q(s,t) &=& \frac{1}{N}\,\sum_{i=1}^{N}\langle\,\eta(s)\,\eta(t)\,\rangle_{(i)}\\
\nonumber G(s,t) &=& \frac{1}{N}\,\sum_{i=1}^{N}\langle\,{\rm i}\widehat{u}_{s}\,\eta(t)\,\rangle_{(i)}
\end{eqnarray}
The $\langle(\ldots)\rangle_{(i)}$ represent averages over the dynamics of effective single site processes and have the form
\begin{widetext}
\begin{equation}
\label{eq:site_i_avg}
\langle(\ldots)\rangle_{(i)} = \frac{\sum_{\{\eta(t)\}} P(\eta(0)) \left( \prod_{t=0}^{T-1}\left[  \int \frac{{\rm d}u(t)\,{\rm d}\hat{u}(t)}{2\pi}\,\delta_{[\eta(t+1),f_i(t)]} \right]\,(\ldots)\, e^{-{\cal{S}}}\right)}{\sum_{\{\eta(t)\}} P(\eta(0)) \left( \prod_{t=0}^{T-1}\left[  \int \frac{{\rm d}u(t)\,{\rm d}\hat{u}(t)}{2\pi}\,\delta_{[\eta(t+1),f_i(t)]} \right]\, e^{-{\cal{S}}}\right)}\,
\end{equation}
\end{widetext}

Averages involving a conjugate field ${\rm i}\widehat{u}(t)$ describe response functions, i.e., perturbations of expectations with respect to an external field. Thus, averages involving only conjugate fields describe a perturbation of a constant and are set to zero. Furthermore, by causality, $G(t,s)$, which describes the response of the fraction of failed processes at time $t$, to a perturbation at time $s$, vanishes $\forall\,s \geq t$. Using the saddle point identities for other order parameters and their conjugates, we obtain  $\Xi_1\,=\,0$ and $\Xi_2\,=\,0$ at the saddle point. The dynamic action reduces to
\begin{eqnarray}
\label{eq:dynamic_action_after_saddle_point_integration}
\nonumber {\cal{S}} &=& \sum_{t=0}^{T-1} \Bigg( {\rm i}\widehat{u}(t)\Big[u(t)\,-\,J_0\,m(t)\\
\nonumber & &-\,\alpha J^2\sum_{s<t}G(t,s)\eta(s)\Big] \Bigg)\,\\
& &-\, \frac{1}{2}\sum_{s,t=0}^{T-1}{\rm i}\widehat{u}(t)\,{\rm i}\widehat{u}(s)\left(J^2\,q(s,t)\,+ \delta_{[s,t]} \right)\,,
\end{eqnarray}
and it corresponds to single site processes, with dynamics
\begin{eqnarray}
\label{eq:effective single site process}
\nonumber \eta(t+1) &=& \Theta \Bigg( J_0\,m(t)\,+\, \alpha\,J^2\,\sum_{s<t}G(t,s)\,\eta(s)\\
& & -\, \vartheta\, +\, \phi(t)\,+\,h(t) \Bigg)\,.
\end{eqnarray}
We note the following: (i) There is now a dependence on the fraction $m(t)$ of failed processes. (ii) If there is a degree of symmetry, $\alpha \neq 0$, the dynamics is non-Markovian, with memory given by the response function $G(t,s)$. (iii) The noise  $\phi(t)$ is colored, with $\langle\,\phi(t)\,\rangle = 0$  and $\langle\,\phi(s)\,\phi(t)\,\rangle = J^2\,q(s,t)\,+ \delta_{[s,t]}$.

We recall that the only $i$ dependence in Eqs. (\ref{eq:resolved observables})-(\ref{eq:site_i_avg}) comes from $\vartheta_i$ in $f_i(t)$, Eq. (\ref{eq:f i t with u i t}). By the law of large numbers the empirical averages in Eq. (\ref{eq:resolved observables}) can be evaluated as an average over the $\vartheta$ distribution,
\begin{equation}
\nonumber
\lim_{N\,\to\,\infty} \frac{1}{N}\sum_{i} \langle(\ldots) \rangle_{(i)}\,=\, \int {\rm d}\vartheta\,p(\vartheta) \langle (\ldots) \rangle \,=\,\overline{ \langle (\ldots) \rangle }^{\vartheta}
\end{equation}
The saddle point results, Eq. (\ref{eq:resolved observables}) thus take the form
\begin{eqnarray}
\label{eq:order para after saddle point}
m(t)\,&=&\, \overline{ \langle \eta(t) \rangle }^{\vartheta}\,,\\
q(s,t) &=& \overline{ \langle \eta(s)\,\eta(t) \rangle  }^{\vartheta}\,,\\ 
G(t,s) &=& \frac{\partial\,m(t)}{\partial\,h(s)}\,.
\end{eqnarray}

In the case $\alpha\,=\,0$, there is no memory effect. Eq. (\ref{eq:order para after saddle point}) is by itself sufficient to describe the dynamics. Evaluating the $\phi(t)$ average we get
\begin{equation}
\label{eq:dynamic observable no sym}
m(t+1)\,=\,\overline{ \Phi \left( \frac{J_0\,m(t)\,-\,\vartheta}{\sqrt{1\,+\,J^2\,m(t)}} \right)}^{\vartheta}\,.
\end{equation}

However, for $\alpha \,\neq\,0$, memory $G(t,s)$ {\it is} relevant. The non-linear nature of the evolution Eqs. (\ref{eq:effective single site process})-(\ref{eq:order para after saddle point}) precludes a simple analytical characterization of the long-time asymptotic stationary states.

\section{Equilibrium Statistical Mechanics}
\label{apx: Replica}
For fully symmetric networks, $\alpha=1$, and thermal noise, $\xi$, distributed as
\begin{equation}
\label{eq: thermal noise2}
p(\xi)\,=\,\frac{1}{2}\,\beta\, {\rm sech}^2 \left( \frac{\beta\,\xi}{2} \right)\,,
\end{equation}
equilibrium statistical mechanics can be employed to study the stationary behavior. The resulting Gibbs-Boltzmann equilibrium distribution for observing a given microscopic state is characterized by the Hamiltonian
\begin{equation}
\label{eq:gibbs boltz dist Hamiltonian}
H(\boldsymbol{\eta})\,=\,-\sum_{i < j}c_{ij}\,J_{ij}\,\eta_{i}\,\eta_{j}\, + \,\sum_{i}\vartheta_{i}\,\eta_{i}\,.
\end{equation}

We obtain the stationary macroscopic order parameter, $m$, describing the fraction of failed processes from the free energy per process. In the limit, $N\,\to\,\infty$, the free energy is expected to be self averaging over the disorder,
\begin{equation}
\label{eq:free energy}
f\,=\,\lim_{N\,\to\,\infty}-\frac{T}{N}\,\overline{\log(Z)}^{c_{ij},x_{ij}}\,.
\end{equation}
The above quenched average is calculated using the ``replica trick",
\begin{equation}
\label{eq:Replica_Trick}
f\, =\,\lim_{N\,\to\,\infty}\,\lim_{n\,\to\,0}  -\dfrac{T}{N\,n}\log{\overline{\,Z^{n}\,}^{c_{ij},x_{ij}}}\,.
\end{equation}

In a manner similar to that presented in appendix \ref{sec:GFA disorder avg}, we take the average over $c_{ij}$ and $x_{ij}$. The resulting expression for the free energy is,
\begin{eqnarray}
\label{eq:free energy after disorder avg}
\nonumber f &=& \lim_{n \to 0}\,\lim_{N \to \infty} -\frac{T}{n\,N}\log \sum_{\{\boldsymbol{\eta}^{\mu}\}}\,\\
\nonumber & &\exp \Bigg(N\,\Bigg( \frac{\beta\,J_{0}}{2}\sum_{\mu}m_{\mu}^{2} +\, \frac{\beta^{2}\,J^{2}}{4} \sum_{\mu,\,\nu}q_{\mu\,\nu}^{2}\,\Bigg)\,\\
& &-\,\beta\sum_{\mu}\sum_{i}\vartheta_{i}\,\eta_{i}^{\mu}\Bigg)\,,
\end{eqnarray}
where,
\begin{equation}
\label{eq: decoupling replica terms}
m_{\mu} = \frac{1}{N}\sum_{i\,=1}^{N}\,\eta_{i}^{\mu}\,, \qquad q_{\mu\, \nu} = \frac{1}{N}\sum_{i\,=\,1}^{N}\,\eta_{i}^{\mu}\,\eta_{i}^{\nu}\,.
\end{equation}
The indices $ 1 \leq \mu, \nu \leq n$ label the replicas. The above parameters were introduced to achieve site factorization, in a manner identical to that in appendix \ref{sec:GFA disorder avg}. Subsequently, the free energy is expressed as an integral, which is evaluated via the saddle point technique.
\begin{widetext}
\begin{eqnarray}
f &=& \lim_{n \to 0}\,\lim_{N \to \infty}\, - \frac{T}{nN}\,\log \int\,{\cal D}(\ldots) \exp\left(N\left[\frac{\beta\,J_{0}}{2}\sum_{\alpha}m_{\alpha}^{2}\,+\,\frac{\beta^{2}\,J^{2}}{4}\sum_{\alpha,\,\epsilon}q_{\alpha \epsilon}^{2}\,+\, {\rm i}\sum_{\alpha}\widehat{m}_{\alpha}\,m_{\alpha}\,+\,{\rm i}\sum_{\alpha,\,\epsilon}\,\widehat{q}_{\alpha \epsilon}\,q_{\alpha \epsilon}\right.\right. \\
\nonumber & &\left. \left. +\,\overline{\,\log \sum_{\{\eta^{\mu}\}}\, \exp \left( -\beta\,H_{{\rm eff}} \right) }^{\vartheta} \right] \right)\,.
\end{eqnarray}
\end{widetext}
Here, the effective Hamiltonian is given by
\begin{eqnarray}
\label{eq: effective hamiltonian}
\nonumber -\beta\,H_{{\rm eff}} &=& -\beta\,\vartheta\sum_{\alpha}\eta^{\alpha}\,-\,{\rm i}\sum_{\alpha}\widehat{m}_{\alpha}\,\eta^{\alpha}\\
& &-\,{\rm i}\sum_{\alpha,\,\epsilon}\widehat{q}_{\alpha \epsilon}\,\eta^{\alpha}\,\eta^{\epsilon}\,.
\end{eqnarray}

\subsection{Saddle Point Equations}
We obtain the following expressions for the order parameters,
\begin{eqnarray}
\label{eq:Order_para_m_and_q} 
m_{\alpha} =  \overline{ \langle \eta^{\alpha} \rangle }^{\vartheta}\,, \qquad
q_{\alpha \epsilon} \,=\, \overline{ \langle \eta^{\alpha} \, \eta^{\epsilon} \rangle }^{\vartheta}\,,
\end{eqnarray}
where,
\begin{equation}
\label{eq:Theta_measure}
\langle (\ldots) \rangle \,=\, \frac{\sum_{\{\eta^{\mu}\}}\,(\ldots)\,\exp \left(-\beta\,H_{{\rm eff}}\right)}{\sum_{\{\eta^{\mu}\}}\, \exp \left(-\beta\,H_{{\rm eff}}\right)}\,.
\end{equation}

A solution of the order parameters is achieved with the following replica symmetric ansatz,
\begin{eqnarray}
\label{eq:RS_Ansatz}
m_{\alpha}\,&=&\,m,\,\,\forall\,\alpha\\
q_{\alpha \epsilon}\,&=&\, m\,\delta_{[\alpha\,,\,\epsilon]}\,+\,q\,(1\,-\,\delta_{[\alpha\,,\,\epsilon]}),\,\,\forall\,\alpha,\,\epsilon
\end{eqnarray}

On application of the ansatz and Gaussian linearization of quadratic replica terms, we obtain
\begin{eqnarray}
\label{eq:m from replica}
m &=& \overline{\int{\cal D}z\,\Phi_{\beta}\, \left(h_{{\rm RS}}\right)}^{\vartheta}\,,\\
\label{eq:q from replica}
q &=& \,\overline{\int{\cal D}z\,\Phi_{\beta}^2\, \left(h_{{\rm RS}}\right)}^{\vartheta}\,,
\end{eqnarray}
where, $\Phi_{\beta}(x) = \frac{1}{2}\left(1\,+\,\tanh\left(\frac{\beta\,x}{2}\right)\right)$ and
\begin{equation}
\label{eq:replicated local field 02}
h_{{\rm RS}} = -\vartheta\,+\,J_{0}\,m\,+\,J\,\sqrt{q}\,z\,+\,\frac{\beta\,J^{2}}{2}\,(m-q)\,.
\end{equation}

\newpage 
\bibliography{draft}

\end{document}